\documentclass[pra,aps,twocolumn,superscriptaddress,showpacs]{revtex4}
\usepackage{amsmath,amsfonts,amssymb}
\usepackage{graphicx}
\usepackage{epsfig}

\begin{document}

\title{Shapiro-like Resonance in Ultracold Molecule Production via an
Oscillating Magnetic Field}
\author{Bin Liu}
\affiliation{Institute of Applied Physics and Computational Mathematics, Beijing 100088,
China}
\affiliation{Graduate School, China Academy of Engineering Physics, Beijing 100088, P. R.
China}
\author{Li-Bin Fu}
\affiliation{Institute of Applied Physics and Computational Mathematics, Beijing 100088,
China}
\affiliation{Center for Applied Physics and Technology, Peking University, 100084,
Beijing, P. R. China}
\author{Jie Liu}
\email{liu_jie@iapcm.ac.cn}
\affiliation{Institute of Applied Physics and Computational Mathematics, Beijing 100088,
China}
\affiliation{Center for Applied Physics and Technology, Peking University, 100084,
Beijing, P. R. China}

\begin{abstract}
We study the process of production of ultracold molecules from ultracold
atoms using a sinusoidally oscillating magnetic field modulation. When the
magnetic field is resonant roughly with the molecular binding energy,
Shapiro-like resonances are observed. Their resonance profiles are well
fitted by the Lorentzian functions. The line widths depend on both the
amplitude and the duration of the applied modulations, and are found to be
dramatically broadened by thermal dephasing effect. The resonance centers
shift due to both many-body effect and finite temperature effect. Our theory
is consistent with recent experiment (S. T. Thompson, E. Hodby, and C. E.
Wieman, Phys. Rev. Lett. \textbf{95}, 190404 (2005)). Our model predicts a
1/3 ceiling for the molecular production yield in uncondensed ultracold
atomic clouds for a long coupling time, while for the condensed atoms the
optimal conversion yield could be beyond the limit.
\end{abstract}

\maketitle

\section{introduction}

Shapiro resonance is one of the most remarkable properties of the
superconducting device, in which, two weakly coupled superconductors are
subject to a voltage difference that is the sum of a dc component $V$ and a
periodic signal $V_m\sin (ft)$. A continuous range of nonzero dc currents
are possible if $V=\frac{\hbar }{2e}kf$, where $2e$ is the Cooper pair
charge, $\hbar $ is the reduced Planck constant, and $k$ is an integer\cite%
{shap63,Barone}. The Shapiro resonance provides a method to measure the
constant of nature $2e/\hbar $ with such precision and universality\cite%
{bloc70,fult73} that, since 1972, the reversed view has been adopted whereby
$2e/\hbar $ is assumed to be known and the above Shapiro resonance is used
to define a standard unit of voltage\cite{Barone,tayl67,pope92}.

Essentially, the Shapiro resonance is a specific phenomenon emerged when the
frequency of the external field is commensurate with the intrinsic frequency
of system. Recently, it has received renewed interests and investigations in
the Bose-Einstein condensates(BEC)\cite{pra59-620,prl95-200401,njp5-94}. For
example, in BEC Josephson junction, the dc value of the drift current shows
up as resonant spikes\cite{pra59-620}. Under experimentally accessible
conditions there exist well-developed half-integer Shapiro-like resonances%
\cite{prl95-200401}. Shapiro effect also allows precise measurements in
atomic BECs. The ac-driven atomic Josephson devices can be used to define a
standard of chemical potential\cite{njp5-94}.

In the present paper, we extend to investigate the Shapiro resonance effects
in ultracold molecule production. The conversion of ultracold atoms to
ultracold molecules by time varying magnetic fields in the vicinity of a
Feshbach resonance is currently a topic of much experimental and theoretical
interest. This particular conversion process lends itself well to the
formation of molecular Bose-Einstein condensates (BECs)\cite%
{Greiner,PRL91-250401,Jochim,PRL92-120401} and atom-molecule superpositions%
\cite{Donley}. These Feshbach molecules and their creation process are also
important for understanding ultracold fermionic systems in the BCS-BEC
crossover regime because they are closely related to the pairing mechanism
in a fermionic superfluid that occurs near a Feshbach resonance\cite%
{Holland,Timmermans,KetterleBCS,GreinerBCS}. We study the process of
production of ultracold molecules from ultracold atoms using a sinusoidally
oscillating magnetic field modulation. The advantage of this method is that
it greatly reduces the heating the cloud experiences in the conversion
process because the conversion occurs far from the center of the Feshbach
resonance. In recent experiments, this technique has been applied and was
shown could produce molecules from atoms more efficiently\cite{PRL95-190404}
and measure the binding energy of Feshbach molecule precisely. However, the
underlying mechanism is not fully understood, the complexity arises from the
many-body problem and time-dependent field involved. The many experimental
observations can not be accounted for by the existing theoretical model. For
example, the Poisson distribution predicted by theory\cite{PRA75-013606} is
departure from the observed Lorentzian-like resonance profiles and how the
finite-temperature effect influences the resonance profiles is still an open
problem. In the present paper, we exploit a microscopic two-channel model to
investigate thoroughly the mechanism underlying the Shapiro resonance
phenomenon in the atom-molecule conversion. With quantitatively considering
the thermal dephasing effect in the uncondensed atom clouds our model could
account for the most experimental observations. Our theory also suggests
some interesting predictions for future's experimental test.

The plan of this paper is as follows. In Sec. II we present our model and
make a thorough analysis on the Shapiro resonance. In Sec. III, with the
inclusion of the dephasing effect in our model, we apply our theory to
explain the recent experiment. In Sec. IV, we extend to discuss the Shapiro
resonance for the case of condensed atoms with emphasizing on the influence
from the interaction between coherent particles. Sec. V is our conclusion.

\section{Shapiro-like resonance in atom-molecule conversion from a
two-channel perspective}

\subsection{Model}

Ignoring the two- and three-body atomic decay and collisional molecular
decay, we exploit the following two-channel microscopic model to describe
the dynamics of converting atoms to molecules in the bosonic system,
\begin{eqnarray}
\hat{H}&= & \left(\epsilon_a-\mu\right)\hat{a}^{\dagger}\hat{a}+
\left(\epsilon_b+\nu(t)-2\mu\right)\hat{b}^{\dagger}\hat{b}  \notag \\
&& +\frac{g}{\sqrt{\mathcal{V}}}\left(\hat{a}^{\dagger}\hat{a} ^{\dagger}%
\hat{b}+\hat{b}^{\dagger}\hat{a} \hat{a}\right).  \label{hambb}
\end{eqnarray}
Here $\hat{a}$ ($\hat{a}^+$) and $\hat{b}$($\hat{b}^+$) are bose
annihilation (creation) operators of atoms and molecules, respectively. The
total number of particles $N=\hat{a}^{\dagger}\hat{a}+2\hat{b}^{\dagger}\hat{%
b}$ is a conserved constant. The atomic and molecular kinetic energies are
given by $\epsilon_a$ and $\epsilon_b$, $\mu$ is the chemical potential, $g$
governs the atom-molecule coupling strength, $\mathcal{V}$ denotes the
quantization volume of trapped particles and therefore $n=N/\mathcal{V}$ is
the mean density of initial bosonic atoms. Where $\nu(t)$ represents the
binding energy of diatomic molecules which depends on the external field,
expressed approximately as\cite{RMP78-1311},
\begin{eqnarray}
\nu(t)=-\frac{\hbar^2}{m (a_{eff}-r_0)^2},  \label{binding}
\end{eqnarray}
where $r_0$ is the effective range of the van der Waals potential, $m$ is
the mass of a bosonic atom, and $a_{eff}$ denotes the effective scattering
length driven by external magnetic field,
\begin{eqnarray}
a_{eff}=a_{bg}\left(1-\frac{\Delta B}{B-B_0}\right),
\end{eqnarray}
where $a_{bg}$ is the background scattering length, $B_0$ is the Feshbach
resonance position, $\Delta B$ is the width of the resonance defined through
the relation with the atom-molecule coupling term $\Delta B=m g^2
/4\pi\hbar^2 |a_{bg} \Delta \mu|$, where $\Delta \mu$ is the difference in
magnetic moment between the closed channel and the open channel state. We
focus on the situation that the selected external field $B_{ex}$ is
modulated sinusoidally with small amplitude $B_{mod}$ and large frequency $%
\omega$ near a Feshbach resonance, i.e.,
\begin{eqnarray}
B(t)=B_{ex}+B_{mod}\sin(\omega t).
\end{eqnarray}
Since $B_{mod}\ll B_{ex}$, the binding energy can be expanded into series to
the first order of $B_{mod}$
\begin{eqnarray}
\nu(t)=\nu_e+\nu_m\sin(\omega t),
\end{eqnarray}
where
\begin{eqnarray}
\nu_e= -\frac{\hbar^2}{m a_{bg}^2}\frac{(B_{ex}-B_0)^2} {\left[\left(1-\frac{%
r_0}{a_{bg}}\right) (B_{ex}-B_0)-\Delta B\right]^2},
\end{eqnarray}
and
\begin{eqnarray}
\nu_m=\frac{\hbar^2}{m a_{bg}^2} \frac{2(B_{ex}-B_0)\Delta B B_{mod}} {\left[%
\left(1-\frac{r_0}{a_{bg}}\right) (B_{ex}-B_0)-\Delta B\right]^3}.
\end{eqnarray}

\subsection{Shapiro-like resonance}

\begin{figure}[t]
\centering
\includegraphics[width=0.5\textwidth]{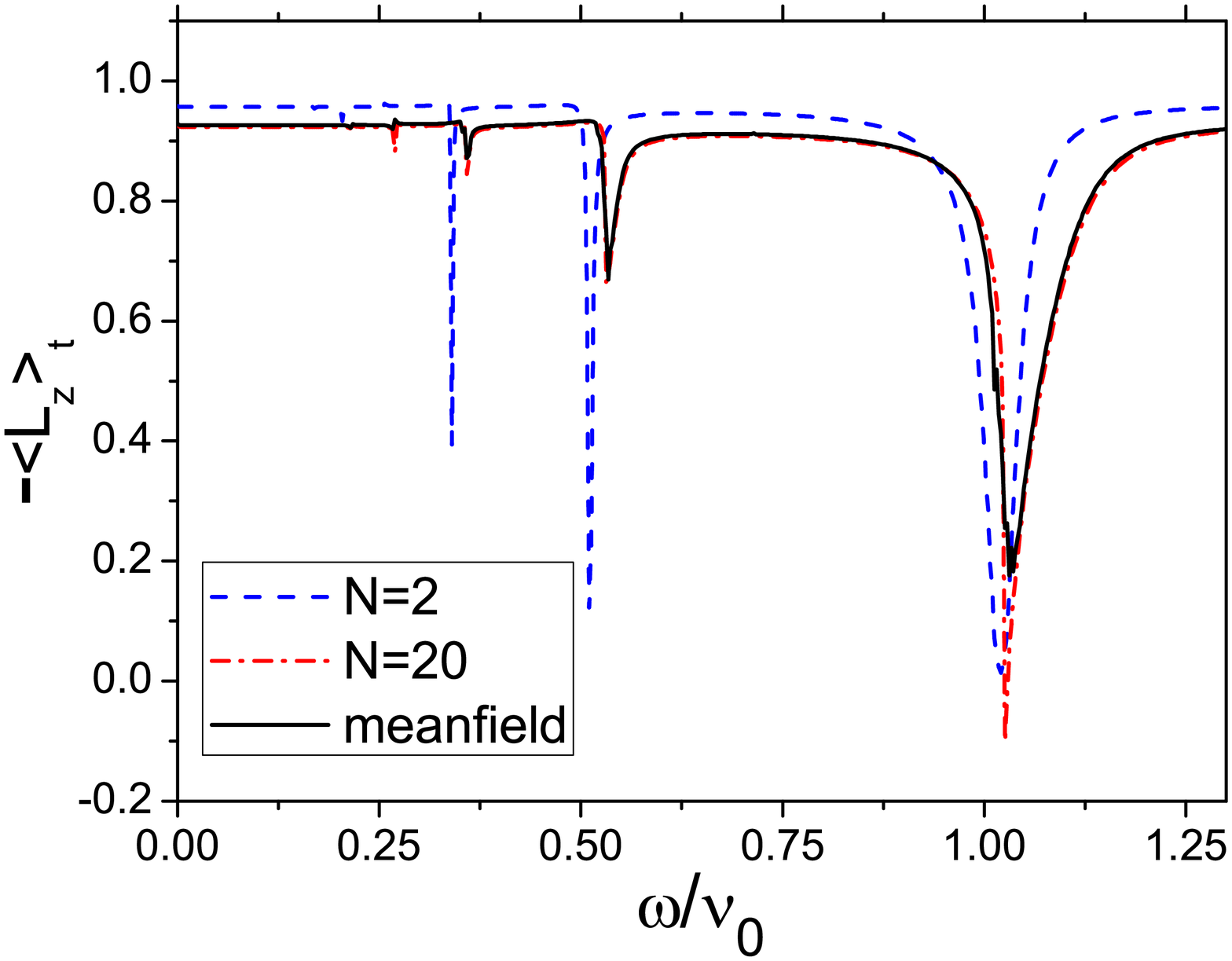}
\caption{Time-averaged population imbalance $-\langle \hat{L}_z \rangle_t$
for the driven system with different numbers of particles, tilt $\protect\nu%
_0/\protect\eta=5$, and scaled driving amplitude $\protect\nu_m=1$.}
\label{fig1}
\end{figure}

We introduce the angular momentum operators to investigate the dynamics of
this system\cite{PRA64-063611},
\begin{eqnarray}
\hat{L}_{x} &=& \sqrt{2}\frac{\hat{a}^{\dagger}\hat{a}^{\dagger}\hat{b}+
\hat{b}^{\dagger}\hat{a} \hat{a}}{N^{3/2}}, \\
\hat{L}_{y} &=& \sqrt{2}i\frac{\hat{a}^{\dagger}\hat{a}^{\dagger}\hat{b}-
\hat{b}^{\dagger}\hat{a}\hat{a}}{N^{3/2}}, \\
\hat{L}_{z} &=& \frac{2\hat{b}^{\dagger}\hat{b}-\hat{a}^{\dagger}\hat{a}}{N},
\end{eqnarray}
with the commutators
\begin{eqnarray}
\left[\hat{L}_z,\hat{L}_x \right]&=& \frac{4i}{N}\hat{L}_y,  \label{lzlx} \\
\left[\hat{L}_z,\hat{L}_y \right]&=& -\frac{4i}{N}\hat{L}_x,  \label{lzly} \\
\left[\hat{L}_x,\hat{L}_y \right] &=& \frac{i}{N}\left(1-\hat{L}%
_z\right)\left(1+3\hat{L}_z\right) +\frac{4i}{N^2}.  \label{lxly}
\end{eqnarray}
Then the Hamiltonian can be written as
\begin{eqnarray}
H=\frac{N}{4}\left((\nu_0+\nu_m\sin(\omega t))\hat{L}_z +\sqrt{2}\eta\hat{L}%
_x\right),
\end{eqnarray}
where $\nu_0=\nu_e+\epsilon_b-2\epsilon_a$ is the energy difference between
atoms and molecules, and parameter $\eta=2g\sqrt{n}$ denotes the coupling
strength. Then the Heisenberg equations of motion are
\begin{eqnarray}
\frac{\mathit{d}}{\mathit{d} t} \hat{L}_x &=& -\frac{1}{\hbar}%
\left(\nu_{0}+\nu_{m}\sin(\omega t)\right) \hat{L}_y,  \label{dlxdt} \\
\frac{\mathit{d}}{\mathit{d} t} \hat{L}_y &=& \frac{1}{\hbar}%
\left(\nu_{0}+\nu_{m}\sin(\omega t)\right) \hat{L}_x -\frac{\eta}{\hbar}%
\frac{\sqrt{2}}{N}  \notag \\
&& +\frac{\eta}{\hbar} \frac{3\sqrt{2}}{4} \left(\hat{L}_z-1\right)\left(%
\hat{L}_z+\frac{1}{3}\right),  \label{dlydt} \\
\frac{\mathit{d}}{\mathit{d} t} \hat{L}_z &=& \frac{\eta}{\hbar} \sqrt{2}
\hat{L}_y.  \label{dlzdt}
\end{eqnarray}
Since $N$ is large for the current experiments and all the commutators
vanish in the limit of $N\rightarrow\infty$, it is appropriate to take $L_x$%
, $L_y$, and $L_z$ as three real numbers $u,v,w$, respectively. Then we get
the mean-field Heisenberg equations
\begin{eqnarray}
\frac{\mathit{d}}{\mathit{d} t} u &=& -\frac{1}{\hbar}\left(\nu_{0}+
\nu_{m}\sin(\omega t)\right) v,  \label{dudt} \\
\frac{\mathit{d}}{\mathit{d} t} v &=& \frac{1}{\hbar}\left(\nu_{0}+
\nu_{m}\sin(\omega t)\right) u  \notag \\
&& +\frac{\eta}{\hbar}\frac{3\sqrt{2}}{4} \left(w-1\right)\left(w+\frac{1}{3}%
\right),  \label{dvdt} \\
\frac{\mathit{d}}{\mathit{d} t} w &=& \frac{\eta}{\hbar}\sqrt{2} v.
\label{dwdt}
\end{eqnarray}
In order to get the time-averaged value of the conversion varied with
different external field, we characterize each quantum trajectory by its
time-averaged imbalance
\begin{eqnarray}
-\langle \hat{L}_z \rangle_t \equiv -\frac{1}{\Delta t} \int_0^{\Delta t} dt
\langle \hat{L}_z \rangle(t),
\end{eqnarray}
employing the averaging interval $\Delta t\gg \hbar/\nu_0$. Initially, all
particles are atoms. Fig.\ref{fig1} shows the results of such calculations
by numerical solving the Heisenberg equations(\ref{dlxdt}-\ref{dlzdt}) for $%
N=2, 20$ under periodic modulation with fixed scaled amplitude $%
\nu_m/\nu_0=0.2$ and frequencies $\omega$ ranging from $0$ to $1.25\nu_0$.
The solution of the mean-field equations (\ref{dudt}-\ref{dwdt}) is also
presented. There are several clear spikes which indicate the Shapiro-like
resonance in atom-molecule conversion driven by external magnetic field.

These spikes indicate that the frequency of the modulated field is
commensurate with the intrinsic frequencies of the atom-molecule conversion
system in the absence of the periodic modulation. Now we analyze the
intrinsic frequency. For $N=2$, using Fock state as basis, the commutators (%
\ref{lzlx}-\ref{lxly}) becomes
\begin{eqnarray}
\left[\hat{L}_z,\hat{L}_x \right]&=& 2i\hat{L}_y, \quad \left[\hat{L}_z,\hat{%
L}_y \right]= -2i\hat{L}_x, \\
\left[\hat{L}_x,\hat{L}_y \right] &=& i\hat{L}_z.
\end{eqnarray}
From the Heisenberg equations(\ref{dlxdt}-\ref{dlzdt}), we get
\begin{eqnarray}
\frac{\mathit{d}^2}{\mathit{d} t^2} \hat{L}_y + \frac{1}{\hbar^2}%
\left(\nu_{0}^2+\eta^2\right) \hat{L}_y =0 .
\end{eqnarray}
Then the intrinsic frequency is readily obtained from the above equation as $%
\sqrt{\nu_0^2+\eta^2}/\hbar$. So the center of resonance is expected to be $%
\sqrt{\nu_0^2+\eta^2}/(\hbar\omega)=p/q$ with $p,q$ are integers. In our
case, the resonances corresponding to $p/q=1,2,3$ are more prominent. With $%
N $ increasing, we find that the resonance center shifts to right due to the
many-body effect. We can obtain the intrinsic frequency in the mean-field
limit,i.e., $N\to \infty$. From the mean-field equations(\ref{dudt}-\ref%
{dwdt}) we readily obtain,
\begin{eqnarray}
\frac{d^2}{dt^2}v+\frac{1}{\hbar^2}\left(\nu_0^2+\eta^2(1-3w)\right)v=0,
\label{freqv}
\end{eqnarray}
Initially all particles are in atom states, i.e., $w=-1$. Approximately
substituting it into the above equation we obtain the explicit expression of
the frequency $\sqrt{\nu_0^2+4\eta^2}/\hbar$. It implies that due to the
many-body effect\cite{PRA78-013618}, the resonance centers shifts to $\sqrt{%
\nu_0^2+4\eta^2}/(\hbar\omega)=1,2,3,...$. The above theoretical analysis
agree with our numerical results.

\subsection{Phase space at the Shapiro-like resonance}

The Shapiro resonance phenomenon can be demonstrated intuitively by the
trajectories in phase space of the system. Notice that the constraint $%
u^{2}+v^{2}=\frac{1}{2}(w-1)^{2}(w+1)$ and introducing the canonical
variable $s=w$, $\theta =\arctan (v/u)$ denoting the population imbalance
and the relative phase between atoms and molecules, the mean-field
Heisenberg equations can be replaced by a classical Hamiltonian of the form
\begin{equation}
\mathcal{H}=\frac{1}{\hbar }\left( \nu _{0}+\nu _{m}\sin (\omega t)\right) s+%
\frac{\eta }{\hbar }\sqrt{(s-1)^{2}(s+1)}\cos \theta ,
\label{claham}
\end{equation}
the canonical equations of motions are
\begin{eqnarray}
\frac{d\theta }{dt} &=&\frac{\partial \mathcal{H}}{\partial s}=\frac{1}{%
\hbar }\left( \nu _{0}+\nu _{m}\sin (\omega t)\right) -\frac{\eta }{\hbar }%
\frac{(1+3s)}{2\sqrt{1+s}}\cos \theta , \\
\frac{ds}{dt} &=&-\frac{\partial \mathcal{H}}{\partial \theta }=\frac{\eta }{%
\hbar }\sqrt{(1-s)^{2}(1+s)}\sin \theta .
\end{eqnarray}%
Using a generation function
\begin{equation}
F(\theta ,S)=\left( \theta -\frac{\nu _{0}t}{\hbar }+\frac{\nu _{m}}{\hbar
\omega }\cos (\omega t)\right) S,
\end{equation}%
with following relations
\begin{eqnarray}
s &=&\frac{\partial F}{\partial \theta }=S, \\
\Theta &=&\frac{\partial F}{\partial S}=\left( \theta -\frac{\nu _{0}t}{%
\hbar }+\frac{\nu _{m}}{\hbar \omega }\cos (\omega t)\right) ,
\end{eqnarray}%
we obtain the new Hamiltonian
\begin{eqnarray}
\mathcal{K}(S,\Theta ) &=&\mathcal{H}+\frac{\partial F}{\partial t}  \notag
\\
&=&\frac{\eta }{\hbar }\sqrt{(1-S)^{2}(1+S)}  \notag \\
&&\cos \left( \Theta +\frac{\nu _{0}t}{\hbar }-\frac{\nu _{m}}{\hbar \omega }%
\cos (\omega t)\right) .
\end{eqnarray}
The secular evolution of $S$ and $\Theta $ can be evaluated from the
time-averaged Hamiltonian,
\begin{equation}
\langle \mathcal{K}(S,\Theta )\rangle _{T}=\frac{1}{T}\int_{0}^{T}\mathcal{K}%
(S,\Theta )dt,
\label{integ}
\end{equation}
with $T=2\pi /\omega .$

\begin{figure}[tb]
\centering
\includegraphics[width=0.5\textwidth]{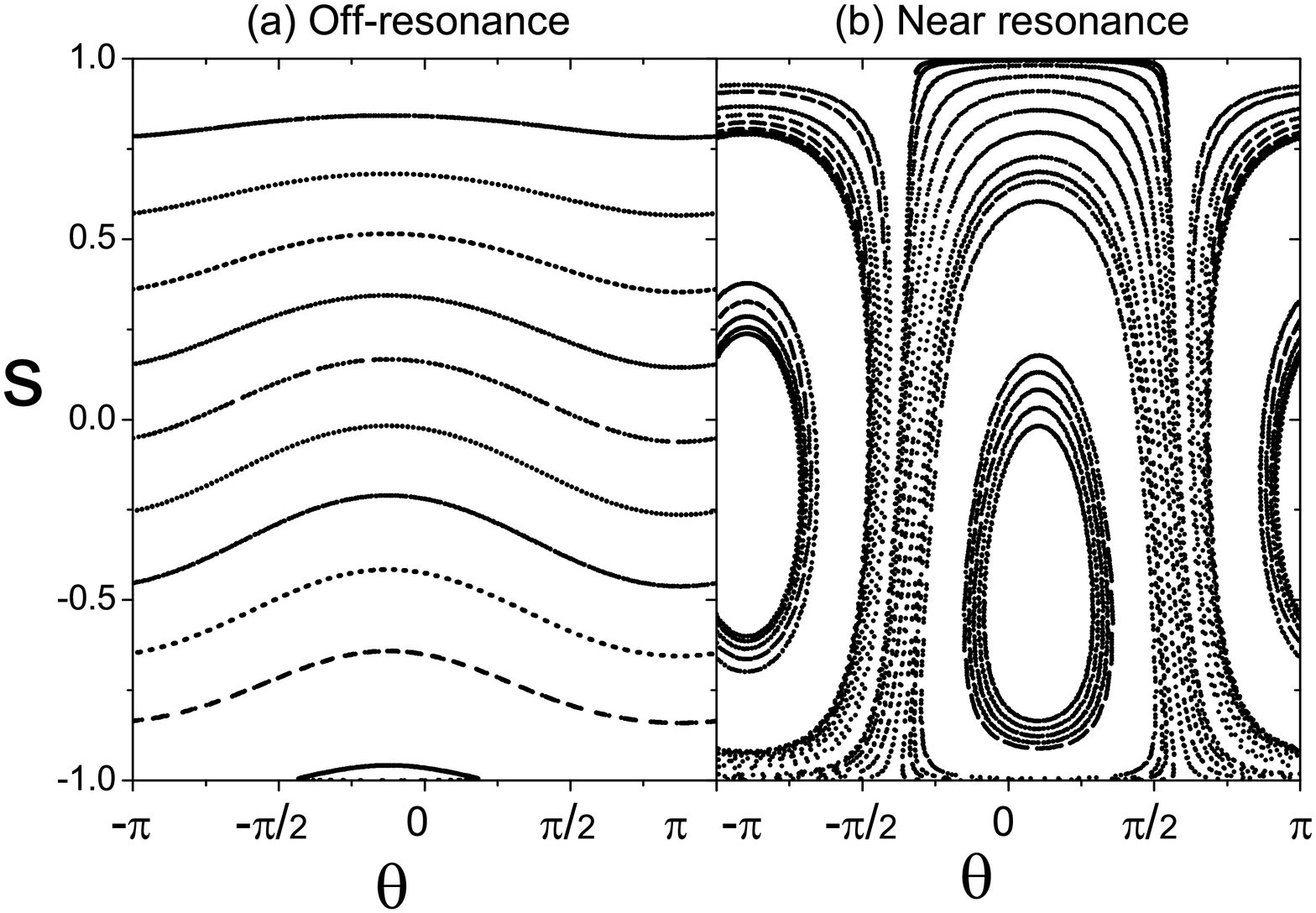}
\caption{Poincar\'{e} section of the classical Hamiltonian (\protect\ref%
{claham}) with tilt $\protect\nu_0/\protect\eta=24$, scaled driving
amplitude $\protect\nu_m=0.2\protect\nu_0$, and modulation frequency (a) $%
\protect\omega/\protect\nu_0=0.95$ (off-resonance), (b) $\protect\omega/%
\protect\nu_0=1$ (near resonance).}
\label{poin}
\end{figure}

Now, we consider the 1:1 resonance case that $\hbar \omega \approx \nu _{0}$%
,
\begin{equation}
\langle \mathcal{K}(S,\Theta )\rangle _{T}\approx J_{1}(\frac{\nu _{m}}{\nu
_{0}})\frac{\eta }{\hbar }\sqrt{(1-S)^{2}(1+S)}\sin \Theta .
\end{equation}
where $J_{1}(x)$ is the first kind Bessel function. The phase graph of the
periodic averaged Hamiltonian system reflects the Poincar\'{e} section of
Hamiltonian system (\ref{claham}), as shown in Fig.\ref{poin}. For the case
of off-resonance, the integral in Eq.(\ref{integ}) approximates to zero, it
implies that the time-averaged $s$ varies a little in time while the
variable $\theta $ increases with time almost linearly. The corresponding
Poincar\'{e} section is shown in Fig.\ref{poin}(b). It is shown that the
phase space at the transition change dramatically.

\subsection{Intrinsic resonance width: Arnold tongues}

For the initial condition $s=w=1$, whether its trajectory falls into
resonance regime can be judged from following resonance condition $\Delta%
\mathcal{H}>2\nu_0/\hbar$, where $\Delta\mathcal{H}$ is the difference
between the maximum and the minimum value of $\mathcal{H}$ in the time
interval $\Delta t\gg \hbar/\nu_0$. For different $\nu_m$ and $\omega$, we
obtain the regions in the two-dimensional parameter space where the
resonance emerges. These regions are named as Arnold tongues\cite{arnold1961}%
. In order to draw out the Arnold tongues in parameter space, the main
numerical tool used in this work is the winding number $\mathcal{W}$
\begin{eqnarray}
\mathcal{W}=\lim_{t\to\infty}\frac{\theta(t)-\theta(0)}{t}.
\end{eqnarray}
with initial conditions $(\theta(0),s(0))$. If the ratio $\mathcal{W}/\omega$
is rational, i.e., $\mathcal{W}/\omega=q/p$, here $q$ and $p$ are natural
number, $(\theta(t),s(t))$ is a resonant solution of $(q:p)$ type, i.e.,
\begin{eqnarray}
\left(\theta(t+pT),s(t+pT)\right)= \left(\theta(t),s(t)\right)+(2\pi q,0),
\end{eqnarray}
\begin{figure}[tb]
\centering
\includegraphics[width=0.5\textwidth]{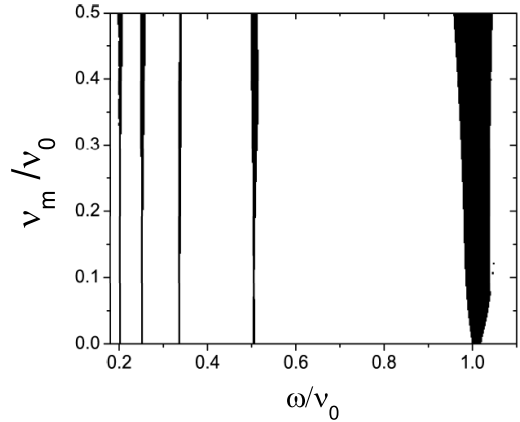}
\caption{Arnold tongues for resonance modes $(1:1),(2:1),(3:1),(4:1),(5:1)$%
(from right to left) with $\protect\nu_0/\protect\eta=5$.}
\label{tongues}
\end{figure}
which means the system runs $q$ times in time $pT$ interval. In our system,
only $(q:1)$ type is significant. In Fig.\ref{tongues} we show the first
five resonance regions with $\nu_0/\eta=5$. The width of the resonance
regions is broadened as the modulation amplitude $\nu_m$ increases.

\section{comparison with experiment}

\begin{figure}[tb]
\centering
\includegraphics[width=0.5\textwidth]{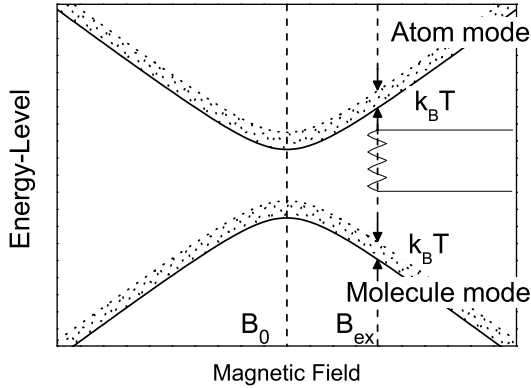}
\caption{Schematic of the swept magnetic field in experiment. See text for
details.}
\label{schematic}
\end{figure}

In the above discussion we use the single-mode model to discuss the
conversion between the condensed atoms and molecules. This is an
approximation because in practical experiments the atoms are not condensed
and other modes will be coupled. We use this approximation under the
condition that the energy distribution of the thermal particles
(characterized by $k_B T$, $k_B$ is the Boltzman constant and $T$ is the
temperature) is much smaller than the effective Feshbach resonance width $g%
\sqrt{n}$\cite{PRA78-013618}. In such cases, each 'energy band' of the
thermal particles can be approximately denoted by one energy level, as
schematically plotted by Fig.\ref{schematic}. Initially, the particles on
one level have a definite phase and the phase difference between two levels
is well defined. However, as the magnetic field sweeps and then the
sinusoidal magnetic field applied, particles will acquire additional phases
that are proportional to their individual energy and evolution time. The
varied particles in one level could acquire different phases because they
have different energies. This define a 'dephasing rate' $\gamma=k_B
T/(2\pi\hbar)$\cite{PRA78-013618}. To to compare with experiment, we need to
include the dephasing effect into our model. Modeling dephasing by fully
include the quantum effects requires sophisticated theoretical studies. The
standard approaches of quantum optics for open systems involve quantum
kinetic master equations. Here, we adopt the simple mean-field treatment in
our model. From the mean-field viewpoint, the decoherence term
introduces a $\gamma$ transversal relaxation term into the mean-field
equations of motion\cite{PRA73-013601},
\begin{eqnarray}
\frac{\mathit{d}}{\mathit{d} t} u &=& -\frac{1}{\hbar}(\nu(t)+\epsilon_b-2%
\epsilon_a) v -\gamma u,  \label{gdudt} \\
\frac{\mathit{d}}{\mathit{d} t} v &=& \frac{1}{\hbar}(\nu(t)+\epsilon_b-2%
\epsilon_a) u  \notag \\
&& +\frac{\eta}{\hbar}\frac{3}{4}\sqrt{2} \left(w-1\right)\left(w+\frac{1}{3}%
\right) -\gamma v,  \label{gdvdt} \\
\frac{\mathit{d}}{\mathit{d} t} w &=& \frac{\eta}{\hbar}\sqrt{2} v,
\label{gdwdt}
\end{eqnarray}
The imbalance of atom-pairs and molecules $w$ is varied in the range of $%
\left[-1,1\right]$ with the lower limit corresponding to a pure atomic gas
and $w=1$ for a pure molecular gas. What we concern is that after the
conversion process how many atomic pairs are converted to molecule. We use $%
w_f$ to denote the value of $w$ when the magnetic field sweep back. The
molecular conversion efficiency can be read from the variable $w_f$ as $%
\Gamma=(1+w_f)/2$.

\begin{figure}[tb]
\centering
\includegraphics[width=0.5\textwidth]{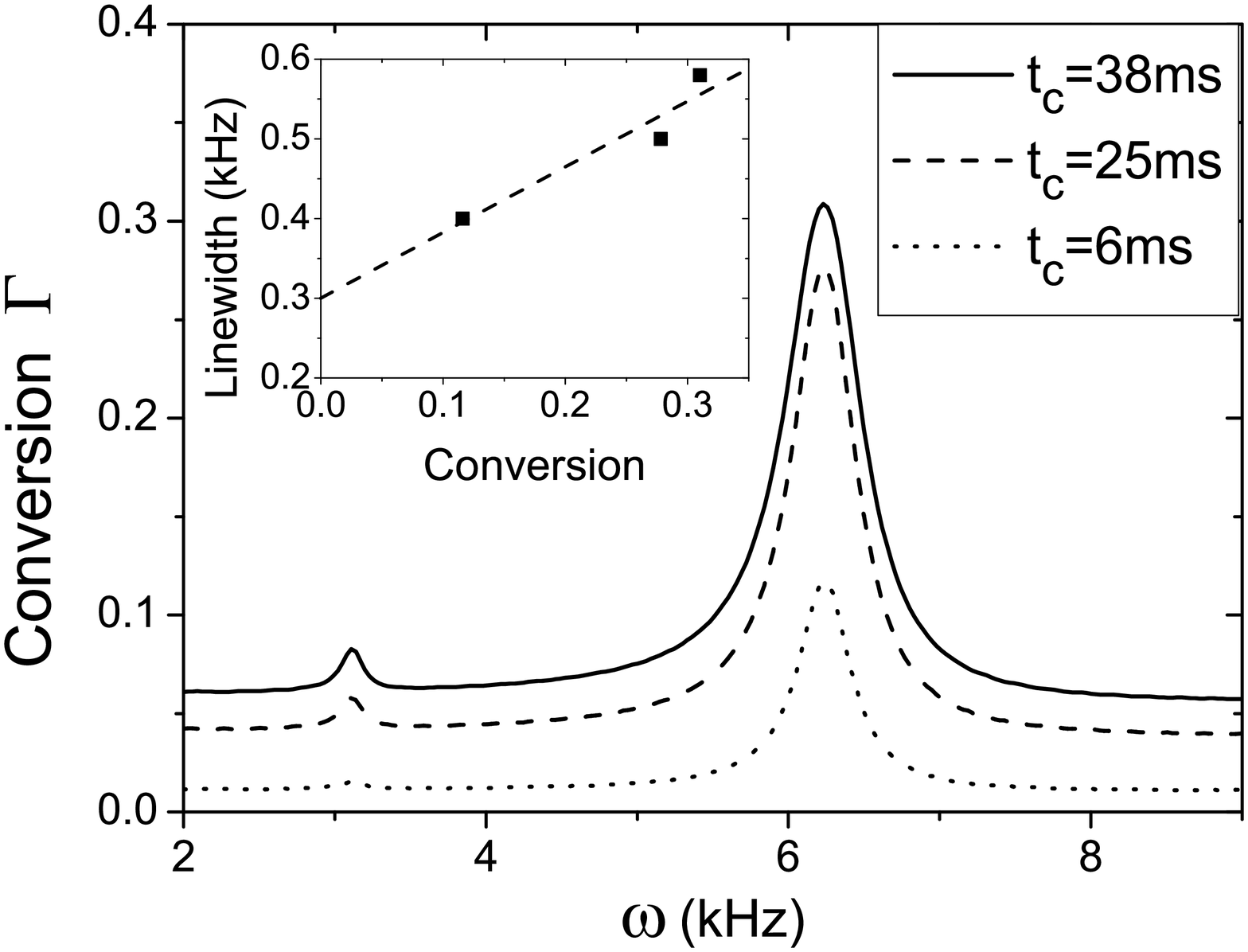}
\caption{Conversion efficiency of atoms converted to molecules as a function
of modulation frequency for three different coupling times. For a fixed
coupling time, the curve can be fitted by a Lorentzian distribution $%
\Gamma=\Gamma_0+\frac{2 A}{\protect\pi}\frac{\Delta}{4(\protect\omega-%
\protect\omega_c)^2+\Delta^2}$, e.g., for $t_c=38ms$, the fitting parameters
are $\Gamma_0=0.06$, $\protect\omega_c=6.2$, $\Delta=0.6$, $A=0.23$. In the
subfigure, by fitting the linewidth versus conversion data to a straight
line we find the zero conversion limit to be $0.3kHz$.}
\label{fig5}
\end{figure}

\begin{figure}[b]
\centering
\includegraphics[width=0.5\textwidth]{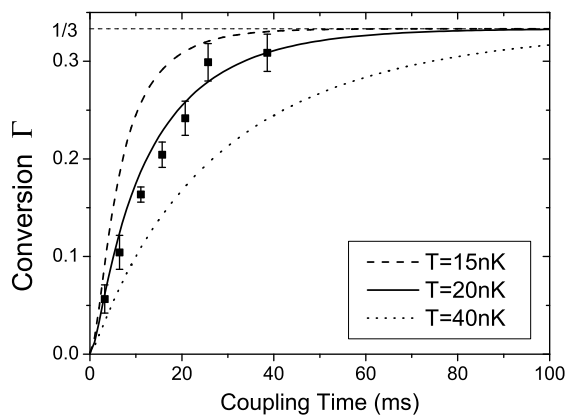}
\caption{Conversion efficiency of atoms converted to molecules under a
periodic modulation with amplitude $B_{mod}=0.13G$ and frequency $\protect%
\omega=6.25kHz$ with respect to coupling time for different temperatures.
The conversion of ultracold atoms to molecules increases with coupling time
until it becomes saturated at $1/3$.}
\label{relaxation}
\end{figure}

Now we apply our theory to the experiment of $^{85}$Rb by Ref.\cite%
{PRL95-190404}. The atoms are held in a purely magnetic trap at a bias field
of $B_{r}$. After evaporative cooling, the magnetic field $B$ is linearly
swept to a selected value at $B_{ex}$ and then apply a sinusoidal magnetic
field pulse with peak-to-peak amplitude $B_{mod}$ and modulation frequency $%
\omega $ for a duration of coupling time. The swept magnetic field can be
expressed as
\begin{equation}
B=\left\{
\begin{array}{ll}
B_{r}-\alpha t & 0\leq t<t_{0} \\
B_{ex}+B_{mod}\sin (\omega t) & t_{0}\leq t<t_{0}+t_{c} \\
B_{ex}+\alpha t & t_{0}+t_{c}\leq t<2t_{0}+t_{c}.
\end{array}
\right.
\label{bfield}
\end{equation}
Where $B_{r}=162G$, $B_{ex}=156.5G$, $B_{mod}=0.13G$, $\omega $ is ranging
from $2kHz$ to $9kHz$, $t_{0}$ is the linear sweep time, $\alpha
=(B_{r}-B_{ex})/t_{0}$ is the linear sweep rate, and $t_{c}$ is the coupling
time. The sketch curve is shown in Fig.\ref{schematic}. For the thermal
cloud, with temperature $T$, one molecule have $5$ degrees of freedom while
two atoms have $6$ degrees of freedom, according to the equipartition
theorem, we have $(2\epsilon _{a}-\epsilon _{b})\approx k_{B}T/2$. The
scaled parameters in Eq.(\ref{gdudt}-\ref{gdwdt}) are
\begin{equation}
\nu (t)=-\frac{\hbar ^{2}}{ma_{bg}^{2}}\frac{(B-B_{0})^{2}}{\left[ \left( 1-%
\frac{r_{0}}{a_{bg}}\right) (B-B_{0})-\Delta B\right] ^{2}},
\end{equation}%
and
\begin{equation}
\eta =2\sqrt{4\pi \hbar ^{2}|a_{bg}\Delta \mu |\Delta Bn/m}.
\end{equation}%
The experimental parameters are $a_{bg}=-443a_{0}$, $r_{0}=185a_{0}$\cite%
{NJP5-69}, $\Delta B=10.71G$, $B_{0}=155G$, $\Delta \mu =1.2\times
10^{-4}\mu _{B}$, the temperature $T=20nK$, density $n=10^{11}cm^{-3}$, here
$a_{0}$ and $\mu _{B}$ are Bohr radius and Bohr magneton, respectively. The
difference of magnetic moment $\Delta \mu $ is extracted from the
experimental data\cite{PRA67-060701}. Under this condition, the ratio
between the energy difference and the energy distribution of thermal
particles, i.e., $\nu _{e}/k_{B}T$, is estimated to be around $15$ that is
much larger than one. The above analysis validates our single mode
approximation. Fig.\ref{fig5} shows the conversion efficiency as a function
of modulation frequency for three different coupling times. The resonance
line width are broadened by the dephasing term. There is a clear Lorentzian
distribution resonance at frequency about $6.25kHz$, close to experiment.
Except the fundamental frequency resonance at $\omega =6.25kHz$, there is
also a weakly $(2:1)$ mode resonance at about $\omega =3.1kHz$, while it has
not been observed in experiment. Our linewidth is approximately $0.3kHz$ at
zero conversion limit, as shown in the subfigure. In the experiment, it is
about $0.2kHz$.

In our calculation, we find that, in the three stages of magnetic field
change expressed by Eq.(\ref{bfield}), the linear process contributes little to the
atom-molecule conversion. This is because oscillation center $B_{ex}$ is
still far away from the Feshbach resonance center. The atom-molecule
conversion mainly occurs in the process of applying the sinusoidal magnetic
field, where $\nu(t)$ can be expressed as $\nu_{0}+ \nu_{m}\sin(\omega t)$.
Therefore, the above observed resonance phenomenon corresponds to the
Shapiro resonance discussed in the last section, while the linewidth is
dramatically broadened by the thermal dephasing effect.

In Fig.\ref{relaxation}, we show the conversion efficiency with respect to
coupling time. For temperature $T=20nK$, density $n=10^{11}cm^{-3}$, our
results are close to experimental data. We also show the cases of different
temperatures by considering the isobaric condition, i.e., $n T=const.$. The
above calculation shows that increasing the temperature will lessen the
molecular production because the dephasing term is proportional to
temperature. On the other aspesct, the conversion efficiency decreases with
the increasing of temperature. For different temperature, a common feature
is that the conversion efficiency increases with coupling time until it
becomes saturated at $1/3$. This can be explained from investigating
Eq.(37-39), where $u=v=0,w=-1/3$ is the fixed point in the absence of the
dephasing term. In the presence of the dephasing effect, the system is
expected to relax into the fixed point in the case of long-term coupling.

\section{Dynamics of atom-molecule conversion for the condensed atoms}

\begin{figure}[tb]
\centering
\includegraphics[width=0.5\textwidth]{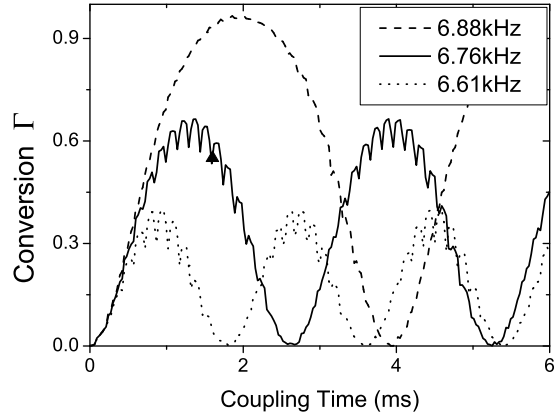}
\caption{The conversion efficiency from condensed atoms to molecules under a
periodic modulation with fixed amplitude $B_{mod}=0.5G$ and different
frequencies . The density for the condensed atoms is $n=10^{12}cm^{-3}$. The
dark triangle marks the experimental observation in Ref.\protect\cite%
{PRL95-190404}.}
\label{conversion}
\end{figure}
\begin{figure}[t]
\centering
\includegraphics[width=0.5\textwidth]{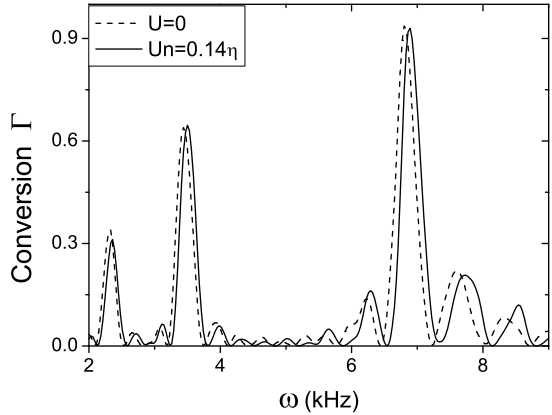}
\caption{The conversion efficiency from condensed atoms to molecules with
respect to modulation frequency for different nonlinear interaction $U$. The
coupling time is $1.6ms$. The other parameters are the same as in Fig.%
\protect\ref{conversion}.}
\label{nonlinear}
\end{figure}

In this section, we extend our discussion to the case of condensed atoms.
For pure BEC atoms, the single mode approximation is valid and dephasing
effect can be ignored while the interaction between the coherent atoms
become significant. After ignoring the kinetic energies of particles, the
Hamiltonian can be written as
\begin{eqnarray}
\hat{H}&= & \nu(t)\hat{b}^{\dagger}\hat{b} -\frac{U}{\mathcal{V}}\hat{a}%
^{\dagger}\hat{a}^{\dagger}\hat{a}\hat{a} +\frac{g}{\sqrt{\mathcal{V}}}\left(%
\hat{a}^{\dagger}\hat{a} ^{\dagger}\hat{b}+\hat{b}^{\dagger}\hat{a} \hat{a}%
\right).  \label{hambb2}
\end{eqnarray}
where $U=4\pi \hbar^2 |a_{bg}|/m$ denotes the nonlinear interaction. In
mean-field limit, we can derive the Heisenberg equations of motion
\begin{eqnarray}
\frac{\mathit{d}}{\mathit{d} t} u &=& -\frac{\nu(t)}{\hbar} v -\frac{2Un}{%
\hbar} v (1-w),  \label{becdu} \\
\frac{\mathit{d}}{\mathit{d} t} v &=& \frac{\nu(t)}{\hbar} u+\frac{\eta}{%
\hbar}\frac{3}{4}\sqrt{2} \left(w-1\right)\left(w+\frac{1}{3}\right)  \notag
\\
&& +\frac{2Un}{\hbar} u (1-w),  \label{becdv} \\
\frac{\mathit{d}}{\mathit{d} t} w &=& \frac{\eta}{\hbar}\sqrt{2} v.
\label{becdw}
\end{eqnarray}

Fig.\ref{conversion} presents the conversion efficiency under a periodic
modulation with fixed amplitude $B_{mod}=0.5G$ and different frequencies by
numerical solving Eq.(\ref{becdu}-\ref{becdw}). The density for the
condensed atoms is $n=10^{12}cm^{-3}$ and thereby the scaled nonlinear
interaction is $Un/\eta=0.14$. One sees that there is a Rabi oscillation
which can reach a high conversion efficiency. In experiment, they observe $%
55\%$ conversion for coupling time $1.6ms$\cite{PRL95-190404}. The
observation marked by a dark triangle in Fig.\ref{conversion} is consistent
with our result. Fig.\ref{nonlinear} presents the conversion efficiency with
respect to modulation frequency for different nonlinear interaction $U$. The
coupling time for this calculation is $1.6ms$. The other parameters are the
same as in Fig.\ref{conversion}.

In Fig.8, we observe some oscillations except for the main resonance peaks
and find that the maximum conversion efficiency could be far beyond the
limit $1/3$. Because the resonance center can be still approximated by $%
\omega=\sqrt{\nu_e^2+4\eta^2}/\hbar$. It implicitly depends on the density
through parameter $\eta$. In the experiment, the density of condensed atoms
is about ten times larger than that of thermal cloud, therefore the
resonance center shift to the right-hand compared to Fig.5 of the thermal
atomic cloud case. Moreover, we find that the interaction between the
coherent atoms will lead to the shift of the resonance profile as clearly
shown in Fig.8.

\section{conclusions}

In conclusion, we have investigated thoroughly the mechanism underlying the
Shapiro resonance phenomenon in the atom-molecule conversion with exploiting
a microscopic two-channel model. With the inclusion of the thermal dephasing
effect in the uncondensed atom clouds our model could account for the most
experimental observations. We also extend our discussions to the case of
condensed atoms. Our theory have some interesting predictions waiting for
future's experimental test.

This work is supported by National Natural Science Foundation of China
(No.10725521,10604009), 973 project of China under Grant No. 2006CB921400,
2007CB814800.

\end{document}